\documentclass[aps,prl,reprint]{revtex4-2}
\usepackage{graphicx}

\usepackage{float}
\usepackage{blindtext}
\usepackage{physics}
\usepackage{gensymb}

\usepackage{comment}

\begin{document}
\title{Atom interferometry with coherent enhancement of Bragg pulse sequences.}

\author{A. B\'eguin}
\author{T. Rodzinka}
\author{L. Calmels}
\author{B. Allard}
\author{A. Gauguet}
\email{gauguet@irsamc.ups-tlse.fr}

\affiliation{Laboratoire Collisions Agr\'egats R\'eactivit\'e, UMR 5589, FERMI, UT3, Universit\'e de Toulouse, CNRS, 118 Route de Narbonne, 31062 Toulouse CEDEX 09, France}

\begin{abstract}
We report here on the realization of light-pulse atom interferometers with Large-momentum-transfer atom optics based on a sequence of Bragg transitions. We demonstrate momentum splitting up to 200 photon recoils in an ultra-cold atom interferometer. We highlight a new mechanism of destructive interference of the losses leading to a sizeable efficiency enhancement of the beam splitters. We perform a comprehensive study of parasitic interferometers due to the inherent multi-port feature of the quasi-Bragg pulses. Finally, we experimentally verify the phase shift enhancement and characterize the interferometer visibility loss.
\end{abstract}

\maketitle
\paragraph{Introduction -}
The demonstrations of matter wave interferometers provided a real breakthrough in the experimental studies of quantum physics, until then limited to thought-experiments. Using their low mass, interferometers based on neutron diffraction have demonstrated interferometers with a large path separation, typically few centimeters \cite{rauch2015}. Neutron interferometry was used for a large number of critical tests in quantum mechanics \cite{rauch2015,Sponar2021}. However, the limited interaction times and the low neutron flux limit the metrological performances in terms of sensitivity. In contrast, ultra-cold atoms allow longer interaction times and remarkable phase sensitivities, but only few experiments have demonstrated large scale atom interferometers \cite{Kovachy2015}. Increasing the separation between the interferometer paths will further improve the sensitivity of inertial sensors \cite{Bongs2019,Geiger2020,Menoret2018,Wu19,Cheiney18}, and fundamental tests such as the equivalence principle \cite{Barret22short,Asenbaum2020,Rosi2017,Zhou15,Tarallo14,Schlippert14,Bonnin13}, quantum electrodynamics \cite{Parker18,Morel20}, and searches for dark matter and energy \cite{Burrage2016,Hamilton2015,El-Neaj2020Short,Sabulsky19,Jaffe2017}. In addition, large scale atom interferometers enable the measurements of Aharanov-Bohm analogues \cite{Gillot13,Gillot14,Overstreet22} and investigate  the fundamental principles of quantum physics and its relation to gravity \cite{overstreet2023b,Marletto2017,Carney2021, Streltsov2022}. Therefore, increasing the number of photon recoils transferred to atoms, with the so-called Large Momentum Transfer beam splitters (LMT), is a major goal in atom interferometry. Various approaches have been demonstrated, most of them create a superposition of two momentum states using a first diffraction pulse followed by an acceleration of each state using continuous Bloch oscillations \cite{Clade2009,McDonald2013,Muller2009,Gebbe2021short} or acceleration based on pulse sequences, either with multi-photon transitions \cite{McGuirk2000,Kotru2015,Chiow2011,Plotkin2018,Jaffe2018}, or single optical photon transitions \cite{Rudolph2020,Wilkason2022}. Each of these solutions has its own advantages that strongly depend on the considered applications and atomic species. So far, atom interferometers with momentum transfer up to 400 photon recoil ($\hbar k$) have been demonstrated with Bloch oscillations in Rb \cite{Gebbe2021short}, and single photon transitions in Sr \cite{Wilkason2022}.
\begin{figure}
\centerline{\includegraphics[width=0.5\textwidth]{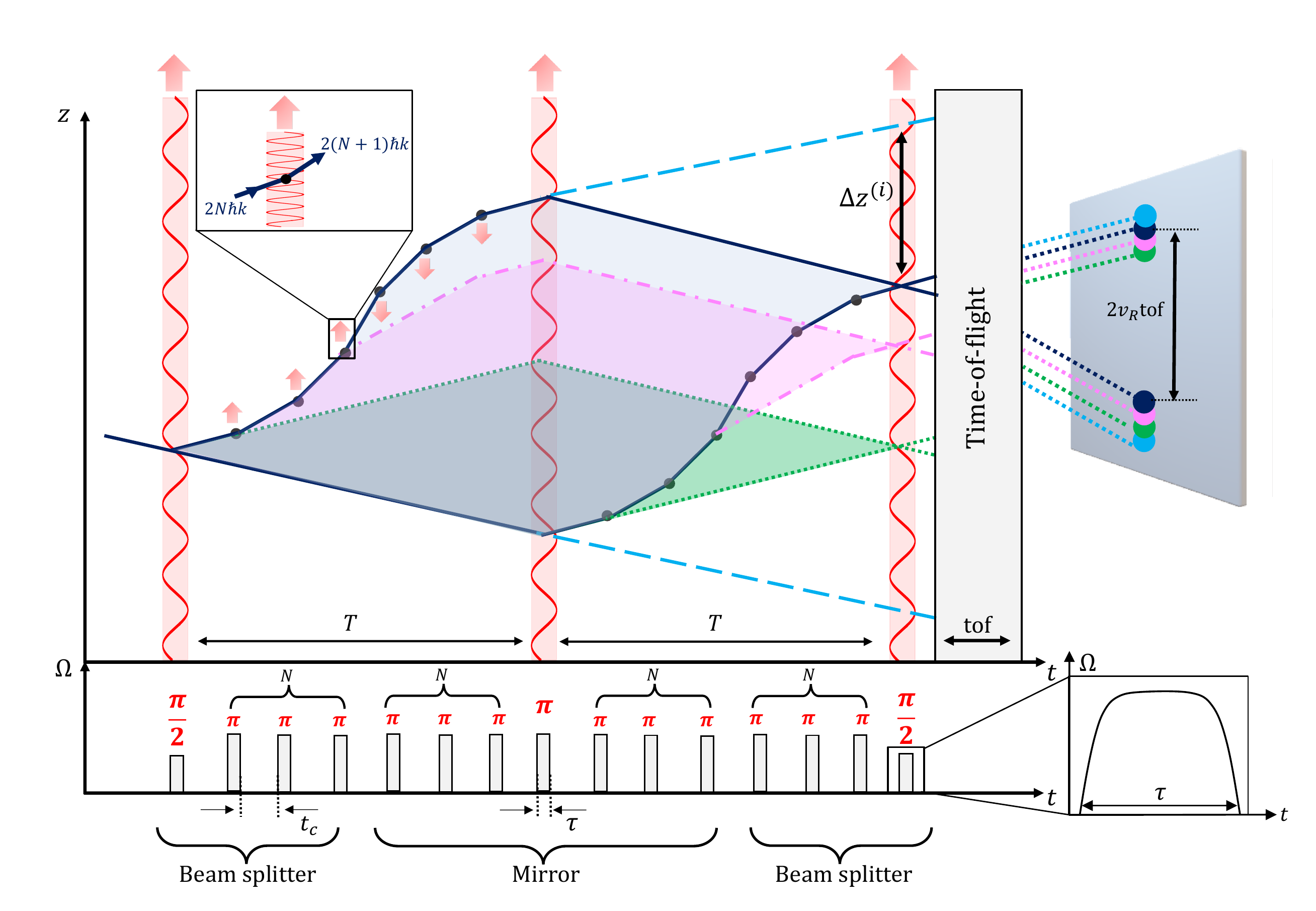}}
\caption{Space-time trajectories for a 8-$\hbar k$-interferometer. The three long red lattices are the diffracting $\pi/2-\pi-\pi/2$ pulses of a standard Mach–Zehnder atom interferometer. The arm separation is increased with sequences of $N$ $\pi$-pulses separated by a time $t_c$, acting only on a single arm (short red arrow). Due to imperfect $\pi$-pulses, loss channels (dashed and dotted lines) can induce parasitic interferometers. The lower panel shows the pulse train of the optical lattice and the hyperbolic tangent amplitude profile used for every pulse.}
\label{fig-LMTsequetial}
\end{figure}

In this letter, we demonstrate phase stable atom interferometers with a maximum momentum separation up to 200$\hbar k$, the largest so far with sequential multi-photon transitions. Other multi-photon methods use smooth amplitude profiles to guarantee the adiabaticity of Bragg transitions \cite{siems20,Muller08b,Beguin2022} and long times between pulses to avoid parasitic interferometers. Here, we implement a rapid sequence of pulses realizing series of diabatic couplings. In this regime, we report an efficiency enhancement based on destructive interferences between loss channels yielding a significant increase of the interferometer visibility. In addition, we study the impact of parasitic interferometers on the visibility estimation. Finally, we demonstrate the phase sensitivity enhancement using LMT interferometer up to 200$\hbar k$.
\begin{figure*}
\centerline{\includegraphics[width=0.95\textwidth]{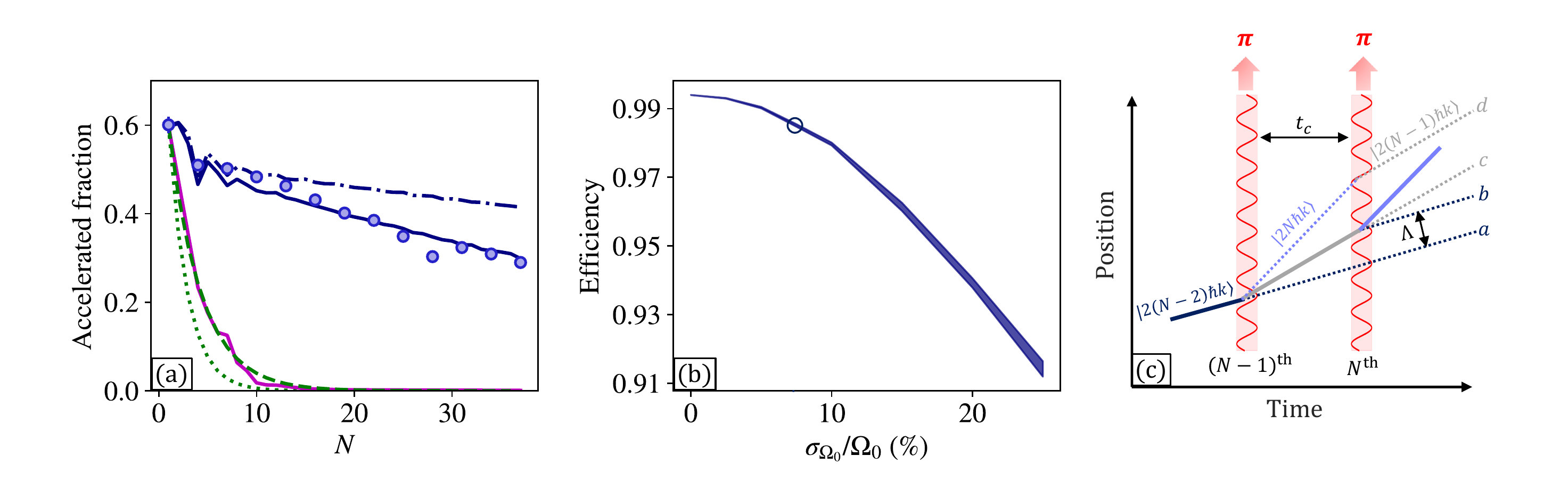}}
\caption{(a) Blue circles: measured atomic fraction in the accelerated state after a sequence of N $\pi$-pulses. Solid dark-blue (resp.  mauve) line is a model including interference between loss channels and pulse-to-pulse fluctuations (resp. without interferences). The dash-dotted line represents the simulated efficiency in the absence of fluctuations. Dotted (resp. dashed) lines are the power law corresponding to independent process for each acceleration pulse for $\sigma_v = 0.3v_r$ (resp. $\sigma_v = 0$). (b) Asymptotic pulse-to-pulse efficiency as a function of the  pulse amplitude standard deviation. The shaded region represents the 1-$\sigma$ statistical uncertainty on the estimation of the efficiency due to stochastic amplitude fluctuations in the optical lattice. The blue circle indicates the value of $\sigma_{\Omega_0}$ reproducing the data (solid dark-blue line in panel (a)). (c) Scheme of the space-time trajectories around the (N-1)-th and N-th acceleration $\pi$-pulses. The solid lines form the accelerated trajectory and the dotted lines are the loss channels. After the 2nd pulse, multiple trajectories (e.g. $a$ and $b$) have the same momentum state.
%(N-1)-th and N-th acceleration $\pi$-pulses, the solid lines are the accelerated paths, the dotted lines the main losses. After the 2nd pulse, outputs a and b correspond to the same momentum state.
These paths interfere destructively if the separation $\Lambda$ between them is smaller than the coherence length and if $t_c \ll \omega_\mathrm{r}^{-1}$.
}
\label{figEfficiency}
\end{figure*}

\paragraph{Experimental setup -} We use a Mach-Zehnder interferometer based on sequences of Bragg pulses (Fig.\ref{fig-LMTsequetial}). The atomic path is split by a beam splitter sequence, the resulting paths are deflected with mirror pulses and the two paths interfere on a second beam splitter sequence. The beam splitter sequence consists in a $\pi/2$-pulse, followed by $N$ acceleration pulses ($\pi$-pulse) driving an almost $100\%$-efficiency momentum transfer at each pulse. The mirror sequence is made of ($2N+1$) $\pi$-pulses which invert the momentum states on each interferometer arm. The pulses are spaced by a time $t_{c}$ and each pulse has the same duration. In order to maximize the number of acceleration pulses, the free evolution time between the end of the beam splitter sequence and the beginning of the mirror sequence is also set to $t_c$. The total interferometer duration $2T + 3\tau$ is then equal to $(4N+2)(\tau+t_{c})+\tau$. The sequences we implement use first order quasi-Bragg diffraction leading to a maximal momentum separation of $2(N+1) \hbar k$ between the interferometer arms. All pulses have a hyperbolic tangent amplitude profile: $\Omega(t) = \Omega_0 \times \max\left\{0,\tanh(\tfrac{8t}{\tau})\tanh(8\left(1-\tfrac{t}{\tau}\right))\right\}$, where $\Omega_{0}$ is the peak two-photon Rabi frequency, proportional to the maximum optical lattice depth. We choose a one-photon detuning $\Delta = 11~\mathrm{GHz}$. At each pulse, the frequency difference of the optical lattice is adjusted to match the Bragg resonance condition in accordance with the atom acceleration. The laser system and optical configuration used to control the optical lattice are described in \cite{Beguin2022}.

An experimental run starts with the preparation of a $^{87}$Rb Bose-Einstein Condensate (BEC) in an optical dipole trap. An horizontal magnetic field gradient during the evaporative cooling polarizes the BEC in the $\ket{F=1,m_F=0}$ state \cite{Cennini03}. The BEC contains approximately $4 \times 10^4$ atoms with an initial velocity spread of $\sigma_v \simeq 0.3 v_\mathrm{r}$ (recoil velocity $v_\mathrm{r}=\hbar k / m$) corresponding to an effective temperature of 50 nK. Prior to the interferometer sequence, the BEC is released in free fall for $4$ ms to acquire an initial non-zero velocity and the first accelerating sequence direction is chosen downward. Following the interferometer sequence, a Time-of-Flight of 15 ms is used to separate the different momentum states. The populations in each state are measured using spatially resolved fluorescence detection (see right part of Fig.\ref{fig-LMTsequetial}). In the data processing, the detection volume is adapted to the sub-recoil temperature of the source. Atoms having undergone a spontaneous emission are partially filtered out of the detection volume. The finite detection region limits the maximum Time-of-Flight to $32$ ms and therefore the number of pulses to about 400 pulses (i.e. a Mach-Zehnder interferometer of $200 \hbar k$) for our typical experimental parameters ($\tau+t_c \approx 30 \; \mu s$).

\paragraph{Coherent Enhancement of Bragg pulses Sequences -}  
Before studying a full interferometer, we present the efficiency of an acceleration sequence. The number of atoms in the fully accelerated state ($\ket{2N\hbar k}$) after a sequence of $N$ $\pi$-pulses is presented in Fig.\ref{figEfficiency}(a). It is normalized to the total detected atom number in all momentum states. The single pulse ($N=1$) efficiency results from a balance between velocity selectivity and non-adiabatic losses. We adjust the lattice parameters ($\Omega_{0} = 7.96 \omega_\mathrm{r} ,\tau = 0.7 \omega_\mathrm{r}^{-1}$), where the recoil frequency $\omega_\mathrm{r} =\hbar k^2/(2m)$, to optimize the efficiency of the N-pulses sequence leading to a single pulse efficiency of $0.6$, which is lower than the optimal single pulse efficiency \cite{Beguin2022}. 

The assumption of independent process for each acceleration pulse would lead to dramatic cumulative losses $0.6^N$, reaching an undetectable remaining fraction for only $N=10$ pulses (green dotted line in Fig.\ref{figEfficiency}(a)). We observe a high efficiency such that more than $30\%$ of the initial atoms are kept in the fully accelerated trajectory after $N=37$ pulses. It corresponds to an asymptotic \textit{pulse-to-pulse efficiency} better than $98\%$. The filtering effect of the velocity distribution during the pulse sequence cannot explain the observed improvement of the diffraction efficiency, since even for zero temperature the number of accelerated atoms would vanish after $N>16$. It is illustrated in Fig.\ref{figEfficiency}(a) by the power law decay of independent pulses at zero velocity dispersion (green dashed line) and by the numerical model for which interference effects are inhibited (mauve solid line). 

The main effect in the efficiency enhancement is due to destructive interferences between the loss channels \cite{Saint-Vincent2010}. This coherent enhancement can be understood using a toy model based on the nearest non-resonant momentum states for each pulse (Fig.\ref{figEfficiency}(c)). The single pulse probability of transfer in these dominant loss channels $\lvert\epsilon \lvert^2$ are of the order of $0.1$, and higher-order paths are less likely. In order to illustrate the interference process, we consider, as an example, the loss channel in the $\ket{2(N-2)\hbar k}$ state at the N-th pulse. Two paths are considered (path $a$ and $b$ Fig.\ref{figEfficiency}(c)): the non-diffracted atoms at the $(N-1)$-th pulse (path a) with the amplitude $\epsilon$ and the non-resonant coupling at the Nth pulse (path b) with the amplitude $\epsilon\exp{i(\pi - 4 \omega_\mathrm{r} t_c)}$. When the coherence length $\xi =\hbar/(m \sigma_v)$ is bigger than the separation between the two paths $\Lambda = 2 v_\mathrm{r} t_c$, they interfere and the population in this loss channel oscillates as:
\begin{equation} 
    P_{\ket{N-2}} = 2 \epsilon^2 [1+\cos(\pi - 4 \omega_\mathrm{r} t_c)].
\end{equation}
For short pulse separations $t_c \ll (4 \omega_\mathrm{r})^{-1}\approx 10~\mu$s, the scattering channels interfere destructively. Similar effects occur for the other loss channels leading to a significant decrease of the non-adiabatic losses at small $t_c$. We have pushed further the investigation with simulations including higher order paths, temperature and pulse-to-pulse amplitude fluctuations. We find a good agreement with experimental data (Fig.\ref{figEfficiency}(a)) confirming a favorable destructive interference of the loss channels. Here, the fluctuations are modeled by normally distributed pulse amplitudes $\Omega_0$ centered on the measured mean value. The standard deviation of the distribution $\sigma_{\Omega_0}$ is a free parameter and is adjusted to 7.5\% to reproduce the data (compatible with the measured pulse-to-pulse fluctuations). Figure~\ref{figEfficiency}(b) presents how the pulse-to-pulse efficiency decreases with the amplitude fluctuations. It shows that a pulse-to-pulse efficiency of 99.3\% could be reached without fluctuations even with our BEC source of $\sigma_v\simeq 0.3v_r$ (dash-dot line in Fig.\ref{figEfficiency}(a)). The efficiency can be further improved with a sub-nK source ($\sigma_v < 0.01 v_r$) \cite{Kovachy2015b,Deppner2021} reaching $>99.9\%$. 

The Coherent Enhancement of Bragg pulse Sequences (CEBS) relaxes the usual trade-off between velocity selection and non-adiabatic losses \cite{siems20} due to the interference of the latter. Therefore, it is interesting to use shorter pulses to attenuate the velocity selectivity and favor the non-adiabatic losses that cancel out in the sequence. Furthermore, the CEBS technique allows for shorter pulse trains than many other acceleration methods, leading to larger space-time area in a given interferometer duration. In reference \cite{Decamps19}, we give a quantitative evaluation of the space-time area for a sequencial LMT interferometer using square pulses. For $N \gg 1$ and $\tau/T \ll 1$, the space-time area ($\mathcal{A^{(N)}}$) is determined by:
\begin{equation}
\label{scalefactor}
    \mathcal{A^{(N)}} = \mathcal{A}^{(0)} + 2 k T^2  N\Big[1  - N\frac{\tau+t_c}{T} \Big],
\end{equation}
where $\mathcal{A}^{(0)}$ is %The equation (\ref{scalefactor}) represents
the space-time area of a standard three-pulse interferometer. The LMT enhancement, given by $N(1 - \frac{N(\tau + t_c)}{T}) > 1$, is smaller than N due to the finite acceleration time. Using a rapid sequence with short $t_c$ and $\tau$ maximizes the space-time area. Additionally, the space-time area increases for any value of N since $T > N(t_c+\tau)$. The increase in the momentum transfer rate is related to the destructive interference of non-adiabatic losses in close connection with the concepts of shortcuts to adiabaticity \cite{DGO19} and composite pulses \cite{LEVITT1986}. Therefore pulse sequence shaping based on optimal control protocols has the potential for further improvement in efficiency and robustness.

\begin{figure}
\centerline{\includegraphics[width=0.5\textwidth]{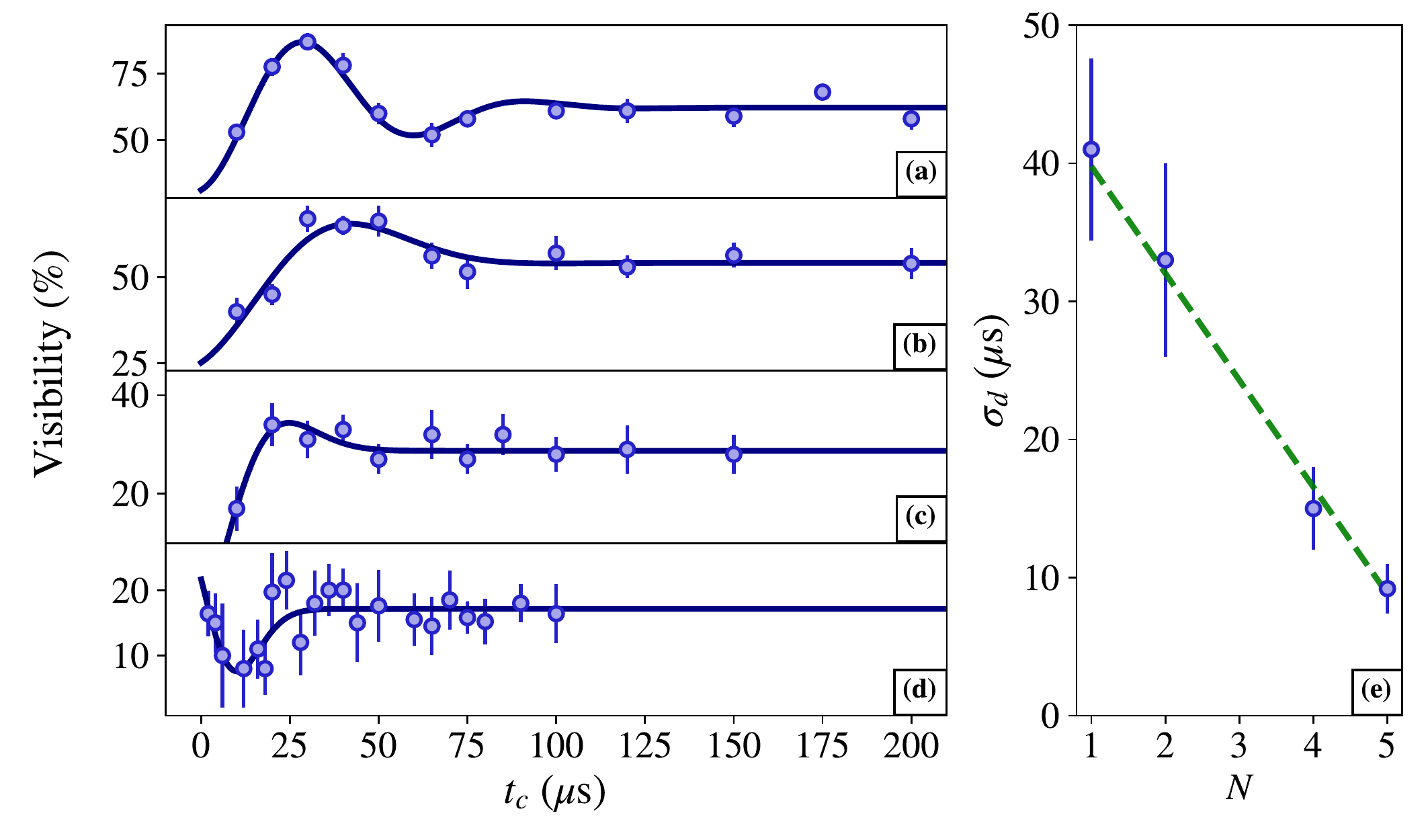}}
\caption{(a-d) Visibility for LMT-interferometer as a function of $t_c$ for $N = [1,2,4,5]$. e) Evolution of the characteristic damping time $\sigma_d$ with N.}
\label{figParasit}
\end{figure}

\paragraph{LMT Interferometer -} We implement LMT Mach-Zehnder interferometers (Fig.\ref{fig-LMTsequetial}) in the CEBS regime and detect the populations in the two main output ports with momentum states $\ket{0}$ and $\ket{2\hbar k}$. Due to imperfect $\pi$-pulses, visibility measurements can be biased by undesired interferometric paths. For short time intervals between pulses ($t_c$), losses into $\ket{0}$ and $\ket{2\hbar k}$ states are counted within the detection volume of the main interferometer. This effect is a consequence of the small separation between the parasitic and main propagation paths ($\Delta z^{(i)}$, as depicted in Fig.~\ref{fig-LMTsequetial}).

In practice, the interferometer visibility $V$ is measured by scanning the phase $\phi_l$ of the optical lattice at the N-th pulse of the first acceleration sequence and by fitting the interference fringes with the function $S = S_0 [1+ V \cos{(\phi_l)}]$. The impact of the spurious paths is therefore strongly reduced: The interferometers resulting from losses at the first pulses, such as the green-dotted-line interferometer in the Fig.\ref{fig-LMTsequetial}, are not scanned. However, there are still parasitic paths that can affect the visibility.

In the following, we distinguish closed and open interferometers. 1) Closed interferometers are very unlikely, since they require one non-resonant transition on each of the two arms in a well defined arrangement (e.g. mauve-dashed-line interferometer in Fig.\ref{fig-LMTsequetial}). In addition, for large $N$, most of these interferometers recombine far from the main interferometer detection region and do not impact the visibility measurements. 2) Parasitic paths create open interferometers with the main paths that modulate the visibility with $t_c$. However, these open interferometers vanish if the distance between path $\Lambda \propto t_c$ is bigger than the coherence length of the atomic ensemble $\Lambda \gg \xi $. 

We measure the visibility of the interference fringes $V$ as a function of $t_c$ for small $N$ values (Fig.\ref{figParasit} (a-d)). The visibility oscillations are damped towards the visibility of the main interferometer $V_0$ for long $t_c$. We determine the characteristic damping time $\sigma_d$ of the parasitic interferometers by fitting the visibility measurements with the function $V = A \exp[-\tfrac{t_c^2}{2\sigma_d^2}]\sin{(\omega t_c + \phi_0)}+V_0$. The parasitic interferometers attenuate more rapidly as N increases (Fig.\ref{figParasit} (e)), which is due to the increase of the distance $\Lambda$ with $N$. We have made an exhaustive inventory of spurious paths correlated with the optical lattice phase jump of $\phi_l$, and estimated the distance $\Lambda$ for each of them. We determine that from $N=20$, these paths are sufficiently separated even for vanishing $t_c$. This is in agreement with the extrapolation of the measurements shown in Fig.\ref{figParasit}(e). We conclude that parasitic interferometers are not a limitation to our visibility measurements in the CEBS regime for large N ($> 20$).

\begin{figure}
\centerline{\includegraphics[width=0.5\textwidth]{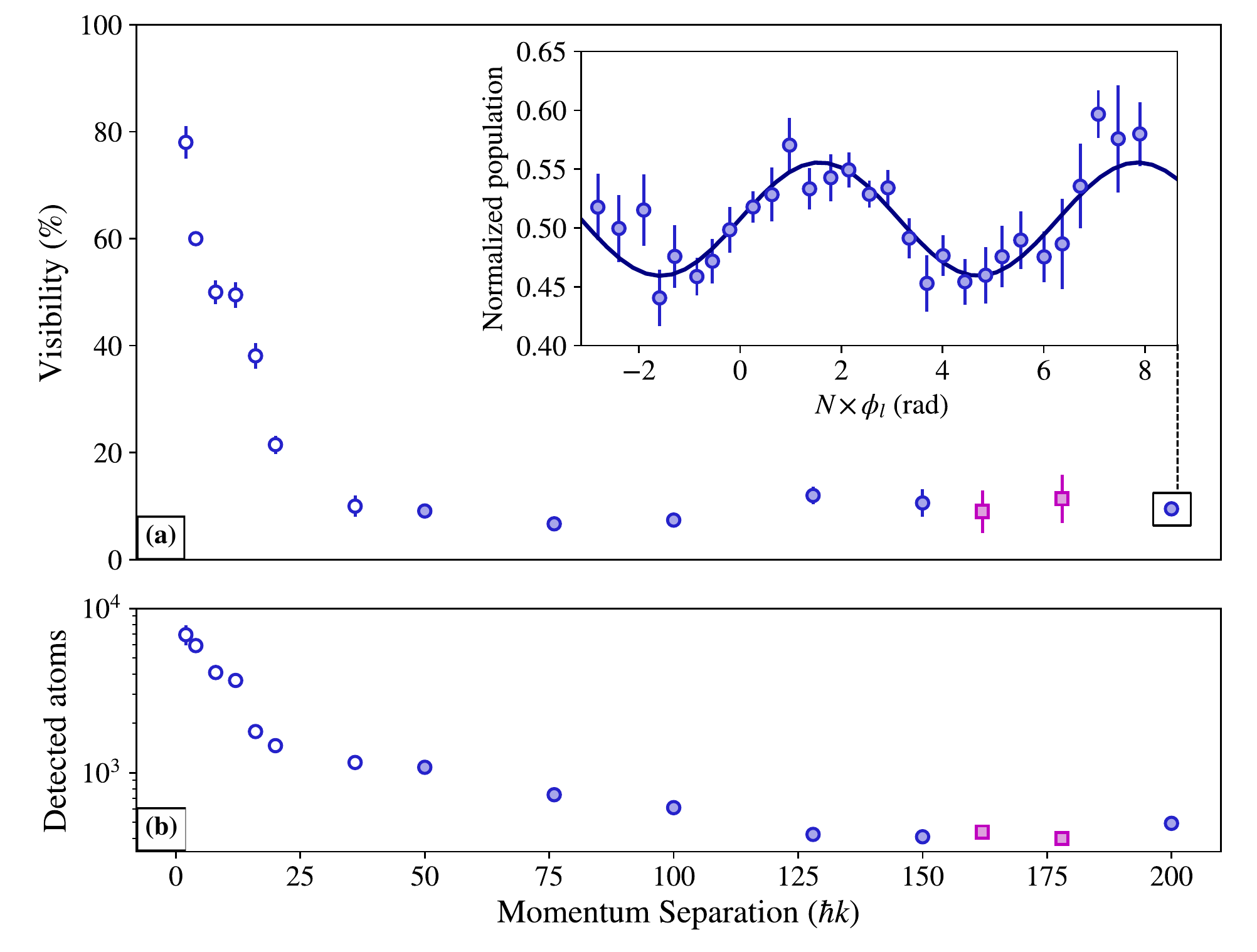}}
\caption{(a) Visibility of LMT-inteferometers as a function of the momentum separation. %($\hbar k_{\mathrm{eff}}$)
For momentum separation $< 34\hbar k$ (open markers) we use long $t_c > 50~\mu$s, while for $\ge 50\hbar k$ (filled markers) we choose short $t_c < 1~\mu$s to benefit from the CEBS. 
The visibility is extracted from a sinusoidal fit of the fringes for a phase jump applied during the last acceleration pulse of the first beam-splitter (mauve squares) or during the first $N$ acceleration pulses (blue circles).
Inset: Fringe pattern for 200$\hbar k$ separation.
(b) Corresponding number of detected atoms.}
\label{figLMT200}
\end{figure}

We present measurements of the interferometer visibilities using a fringe scan on the last acceleration pulse for 162 and 178$\hbar k$ (N = 80 and 88) momentum splitting and $t_c=1~\mu$s, in the CEBS regime. The estimated visibility is $9 \pm 4 \%$ for 162$\hbar k$ and $11 \pm 4 \%$ for 178$\hbar k$ momentum separation (see mauve squares in \ref{figLMT200}(a)).

We verify the scaling of the interferometer phase shift with the laser phase \cite{Decamps19} to further support that the measured visibility is due to the main interferometer and not to parasitic interferometers. These measurements are made with a laser phase jump $\phi_l$ integrated during the first N acceleration pulses (blue dots in Fig.\ref{figLMT200}), which leads to a phase shift $N \times \phi_l$, thus providing a clear signature of the $2(N+1)\hbar k$-interferometer. The inset in Fig.\ref{figLMT200}(a) shows the interferometer fringe for 200$\hbar k$ separation. Both ways of scanning the interference fringes give similar visibility estimations, indicating that the parasitic interferometers do not induce a significant bias.

The visibility measurements and the corresponding number of detected atoms with various LMT-interferometers are shown in Fig.\ref{figLMT200}. For interferometers with momentum separation lower than 34$\hbar k$ (N=17), we use long time between acceleration pulses ($t_c > 50~\mu$s) to strongly suppress parasitic interferometers. In this regime, we do not benefit from the destructive interference between the losses, which explains the rapid drop in the number of detected atoms (Fig.\ref{figLMT200}(b)). In addition, a part of the losses is still detected but does not interfere, resulting in a steep reduction in visibility (Fig.\ref{figLMT200}(a)). For larger LMT-interferometers ($\ge 50 \hbar k$), we use a short time between pulses $t_c =1~\mu$s and take advantage of the CEBS. In this regime, we maintain a visibility up to 200$\hbar k$, measured at $9\pm 1\%$ (inset Fig.~\ref{figLMT200}(a)). This maximum momentum splitting is solely limited by the available Time-of-Flight in our setup. The main candidates for the limited values of the visibility are the laser beam inhomogeneities, and losses in the LMT sequences. The losses are due to spontaneous emission and the finite efficiency of the pulse sequences, which reduce the detected atom number to about $1\%$ of its initial value. These losses can degrade the visibility if a part of these non-interfering atoms is detected in the two main ports. Finally, the laser beam inhomogeneities can cause a dephasing over the atomic ensemble such as AC-Stark Shift \cite{Kovachy2015}, wavefront distortions \cite{Bade2018}, and diffraction phase \cite{Buchner2003,Estey2015,siems2022}, which leads to a visibility reduction after ensemble averaging. Further theoretical studies are required to provide a quantitative understanding of the limits in visibility and detected atoms and to reach the optimal efficiencies. Nevertheless, the slow decay in visibility and detected atom number is very appealing for the scalability of LMT-interferometers based on the CEBS.

\paragraph{Conclusion -} We have implemented LMT-interferometers up to 200$\hbar k$, improving upon the previous record of 112$\hbar k$ based on a sequence of quasi-Bragg pulses \cite{Plotkin2018}. We demonstrate a pulse-to-pulse efficiency $> 98\%$ resulting from destructive interference processes in the loss channels when using diabatic pulses. This Coherent Enhancement of Bragg pulses Sequences (CEBS) leads to a slow decay in visibility, promising a robust LMT scalability. Furthermore, the current limitations can be mitigated using more laser power and AC-Stark shift compensation strategies \cite{Kim20}. In order to illustrate the potential of the CEBS we have performed numerical simulations, indicating that pulse-to-pulse efficiency $> 99.9 \%$ is conceivable, which makes momentum splitting of 1000$\hbar k$ within reach. In addition, CEBS could be further improved in combination with optimal control of the pulse sequence \cite{Goerz2023,louie2023,saywell2023short} or cavity power enhancement \cite{Hamilton15,sabulsky2022}. Finally, the LMT beam splitters reported here rely on short pulse trains allowing faster LMT beam splitters compared to existing multi-photon atom optics. For example, simulations show that a 200$\hbar k$ momentum splitting in less than a millisecond is feasible, while maintaining a high efficiency. Therefore this technique is particularly suitable for quantum sensors based on LMT interferometers with limited Time-of-Flight. Furthermore, the anticipated scalability could contribute to the design of large scale atom interferometers \cite{Abe2021short,Badurina2020short,Canuel2018short,Dimopoulos08,zhan2020short,schlippert} proposed for gravitational wave detectors and exploration of physics beyond the standard model.
 
\begin{acknowledgments}
We thank Jacques Vigu\'e and Naceur Gaaloul for useful discussion and comments during the preparation of this manuscript. We would like to acknowledge the support from French government grants managed by the ANR: TANAI-19-CE47-0002-01 and the "Plan France 2030" ANR-22-PETQ-0005.
\end{acknowledgments}

A. B. and T. R. contributed equally to this work.

\end{document}